# Beyond probability-impact matrices in project risk management: A quantitative methodology for risk prioritisation

F. Acebes [1✉], J. M. González-Varona[2], A. López-Paredes[2] & J. Pajares[1]

The project managers who deal with risk management are often faced with the difficult task of determining the relative importance of the various sources of risk that affect the project. This prioritisation is crucial to direct management efforts to ensure higher project profitability. Risk matrices are widely recognised tools by academics and practitioners in various sectors to assess and rank risks according to their likelihood of occurrence and impact on project objectives. However, the existing literature highlights several limitations to use the risk matrix. In response to the weaknesses of its use, this paper proposes a novel approach for prioritising project risks. Monte Carlo Simulation (MCS) is used to perform a quantitative prioritisation of risks with the simulation software MCSimulRisk. Together with the definition of project activities, the simulation includes the identified risks by modelling their probability and impact on cost and duration. With this novel methodology, a quantitative assessment of the impact of each risk is provided, as measured by the effect that it would have on project duration and its total cost. This allows the differentiation of critical risks according to their impact on project duration, which may differ if cost is taken as a priority objective. This proposal is interesting for project managers because they will, on the one hand, know the absolute impact of each risk on their project duration and cost objectives and, on the other hand, be able to discriminate the impacts of each risk independently on the duration objective and the cost objective.

[1] GIR INSISOC. Dpto. de Organización de Empresas y CIM. Escuela de Ingenierías Industriales, Universidad de Valladolid, Pº Prado de la Magdalena s/n, 47011 Valladolid, Spain. [2] GIR INSISOC. Dpto. Economía y Administración de Empresas, Universidad de Málaga, Avda. Cervantes, 2, 29071 Málaga, Spain. ✉email: fernando.acebes@uva.es

## Introduction

The European Commission (2023) defines a project as a temporary organizational structure designed to produce a unique product or service according to specified constraints, such as time, cost, and quality. As projects are inherently complex, they involve risks that must be effectively managed (Naderpour et al. 2019). However, achieving project objectives can be challenging due to unexpected developments, which often disrupt plans and budgets during project execution and lead to significant additional costs. The Standish Group (2022) notes that managing project uncertainty is of paramount importance, which renders risk management an indispensable discipline. Its primary goal is to identify a project's risk profile and communicate it by enabling informed decision making to mitigate the impact of risks on project objectives, including budget and schedule adherence (Creemers et al. 2014).

Several methodologies and standards include a specific project risk management process (Axelos, 2023; European Commission, 2023; Project Management Institute, 2017; International Project Management Association, 2015; Simon et al. 1997), and there are even specific standards and guidelines for it (Project Management Institute, 2019, 2009; International Organization for Standardization, 2018). Despite the differences in naming each phase or process that forms part of the risk management process, they all integrate risk identification, risk assessment, planning a response to the risk, and implementing this response. Apart from all this, a risk monitoring and control process is included. The "Risk Assessment" process comprises, in turn, risk assessments by qualitative methods and quantitative risk assessments.

A prevalent issue in managing project risks is identifying the significance of different sources of risks to direct future risk management actions and to sustain the project's cost-effectiveness. For many managers busy with problems all over the place, one of the most challenging tasks is to decide which issues to work on first (Ward, 1999) or, in other words, which risks need to be paid more attention to avoid deviations from project objectives.

Given the many sources of risk and the impossibility of comprehensively addressing them, it is natural to prioritise identified risks. This process can be challenging because determining in advance which ones are the most significant factors, and how many risks merit detailed monitoring on an individual basis, can be complicated. Any approach that facilitates this prioritisation task, especially if it is simple, will be welcomed by those willing to use it (Ward, 1999).

Risk matrices emerge as established familiar tools for assessing and ranking risks in many fields and industry sectors (Krisper, 2021; Qazi et al. 2021; Qazi and Simsekler, 2021; Monat and Doremus, 2020; Li et al. 2018). They are now so commonplace that everyone accepts and uses them without questioning them, along with their advantages and disadvantages. Risk matrices use the likelihood and potential impact of risks to inform decision making about prioritising identified risks (Proto et al. 2023). The methods that use the risk matrix confer higher priority to those risks in which the product of their likelihood and impact is the highest.

However, the probability-impact matrix has severe limitations (Goerlandt and Reniers, 2016; Duijm, 2015; Vatanpour et al. 2015; Ball and Watt, 2013; Levine, 2012; Cox, 2008; Cox et al. 2005). The main criticism levelled at this methodology is its failure to consider the complex interrelations between various risks and use precise estimates for probability and impact levels. Since then, increasingly more academics and practitioners are reluctant to resort to risk matrices (Qazi et al. 2021).

Motivated by the drawbacks of using risk matrices or probability-impact matrices, the following research question arises: Is it possible to find a methodology for project risk prioritisation that overcomes the limitations of the current probability-impact matrix?

To answer this question, this paper proposes a methodology based on Monte Carlo Simulation that avoids using the probability-impact matrix and allows us to prioritise project risks by evaluating them quantitatively, and by assessing the impact of risks on project duration and the cost objectives. With the help of the 'MCSimulRisk' simulation software (Acebes et al. 2024; Acebes et al. 2023), this paper determines the impact of each risk on project duration objectives (quantified in time units) and cost objectives (quantified in monetary units). In this way, with the impact of all the risks, it is possible to establish their prioritisation based on their absolute (and not relative) importance for project objectives. The methodology allows quantified results to be obtained for each risk by differentiating between the project duration objective and its cost objective.

With this methodology, it also confers the 'Risk Assessment' process cohesion and meaning. This process forms part of the general Risk Management process and is divided into two subprocesses: qualitative and quantitative risk analyses (Project Management Institute, 2017). Although Monte Carlo simulation is widely used in project risk assessments (Tong et al. 2018; Taroun, 2014), as far as we know, the literature still does not contain references that use the data obtained in a qualitative analysis (data related to the probability and impact of each identified risk) to perform a quantitative risk analysis integrated into the project model. Only one research line by A. Qazi (Qazi et al. 2021; Qazi and Dikmen, 2021; Qazi and Simsekler, 2021) appears, where the authors propose a risk indicator with which they determine the level of each identified risk that concerns the established threshold. Similarly, Krisper (2021) applies the qualitative data of risk factors to construct probability functions, but once again falls in the error of calculating the expected value of the risk for risk prioritisation. In contrast, the novelty proposed in this study incorporates into the project simulation model all the identified risks characterised by their probability and impact values, as well as the set of activities making up the project.

In summary, instead of the traditional risk prioritisation method to qualitatively estimate risk probabilities and impacts, we model probabilities and impacts (duration and cost) at the activity level as distribution functions. When comparing both methods (traditional vs. our proposal), the risk prioritisation results are entirely different and lead to a distinct ranking.

From this point, and to achieve our purpose, the article comes as follows. **Literature review** summarises the relevant literature related to the research. **Methodology** describes the suggested methodology. **Case study** presents the case study used to show how to apply the presented method before discussing the obtained results. Finally, **Conclusions** draws conclusions about the proposed methodology and identifies the research future lines that can be developed from it.

## Literature review

This section presents the literature review on risk management processes and probability-impact matrices to explain where this study fits into existing research. This review allows us to establish the context where our proposal lies in integrated risk management processes. Furthermore, it is necessary to understand the reasons for seeking alternatives to the usual well-known risk matrices.

**Risk management methodologies and standards.** It is interesting to start with the definition of 'Risk' because it is a term that is not universally agreed on, even by different standards and norms. Thus, for example, the International Organization for Standardization (2018) defines it as "the effect of uncertainty on objectives", while the Project Management Institute (2021) defines it as "an uncertain event or condition that, if it occurs, has a positive or negative effect on one or more project objectives". This paper adopts the definition of risk proposed by Hillson (2014), who uses a particular concept: "risk is uncertainty that matters". It matters because it affects project objectives and only the uncertainties that impact the project are considered a 'risk'.

Other authors (Elms, 2004; Frank, 1999) identify two uncertainty categories: aleatoric, characterised by variability and the presence of a wide range of possible values; epistemic, which arises due to ambiguity or lack of complete knowledge. Hillson (2014) classifies uncertainties into four distinct types: aleatoric, due to the reliability of activities; stochastic, recognised as a risk event or a possible future event; epistemic, also due to ambiguity; ontological, that which we do not know (black swan). Except for ontological uncertainty, which cannot be modelled due to absolute ignorance of risk, the other identified uncertainties are incorporated into our project model. For this purpose, the probability and impact of each uncertainty are modelled as distribution functions to be incorporated into Monte Carlo simulation.

A risk management process involves analysing the opportunities and threats that can impact project objectives, followed by planning appropriate actions for each one. This process aims to maximise the likelihood of opportunities occurring and to minimise the likelihood of identified threats materialising.

Although it is true that different authors have proposed their particular way of understanding project risk management (Kerzner, 2022; Hillson and Simon, 2020; Chapman and Ward, 2003; Chapman, 1997), we wish to look at the principal methodologies, norms and standards in project management used by academics and practitioners to observe how they deal with risk (Axelos, 2023; European Commission, 2023; International Organization for Standardization, 2018; Project Management Institute, 2017; International Project Management Association, 2015) (Table 1).

Table 1 shows the main subprocesses making up the overall risk management process from the point of view of each different approach. All the aforementioned approaches contain a subprocess related to risk assessment. Some of these approaches develop the subprocess by dividing it into two parts: qualitative assessment and quantitative assessment. Individual project risks are ranked for further analyses or action with a qualitative assessment by evaluating the probability of their occurrence and potential impact. A quantitative assessment involves performing a numerical analysis of the joint effect of the identified individual risks and additional sources of uncertainty on the overall project objectives (Project Management Institute, 2017). In turn, all these approaches propose the probability-impact or risk matrix as a technique or tool for prioritising project risks.

Within this framework, a ranking of risks by a quantitative approach applies as opposed to the qualitative assessment provided by the risk matrix. To do so, we use estimates of the probability and impact associated with each identified risk. The project model includes these estimates to determine the absolute value of the impact of each risk on time and cost objectives.

**Probability-impact matrix**. The risk matrix, or probability-impact matrix, is a tool included in the qualitative analysis for risk management and used to analyse, visualise and prioritise risks to make decisions on the resources to be employed to combat them (Goerlandt and Reniers, 2016; Duijm, 2015). Its well-established

**Table 1 Risk Management Processes.**

| | MANAGEMENT | IDENTIFICATION | ASSESSMENT | PLANNING RESPONSE | RESPONSE IMPLEMENT. | MONITORING AND CONTROL |
|---|---|---|---|---|---|---|
| **ISO 31000** | 1. Communication and consultation<br>2. Scope, context, criteria | 3.1. Risk identification | ***3.2. Risk analysis***<br>***3.3. Risk evaluation*** | 4. Risk treatment | | 5. Monitoring and review<br>6. Recording and reporting |
| **PMBoK** | 1. Plan risk management | 2. Identify risks | ***3. Qualitative risk analysis***<br>***4. Quantitative risk analysis*** | 5. Plan risk response | 6. Implement risk response | 7. Monitoring and controlling risks |
| **Prince2** | 1.1. Identify the context | 1.2. Identify risks | ***2.1. Risk estimation***<br>***2.2. Risk assessment*** | 3. Plan response to risks | 4. Implement response to risks | 5. Communicate |
| **PM²** | | 1. Identify risks | ***2. Risk assessment*** | 3. Develop risk response | | 4. Risk Control |
| **ICB 4.0** | 1. Risk management framework | 2. Identify opportunities and threats | ***3. Risk assessment*** | 4. Select risk response | | 5. Control |

Norms, standards and methodologies. Source: Rehacek (2017).

| Likelihood \ Impact | Negligible | Minor | Moderate | Significant | Severe |
|---|---|---|---|---|---|
| Very Likely | Low Med | Medium | Med Hi | High | High |
| Likely | Low | Low Med | Medium | Med Hi | High |
| Possible | Low | Low Med | Medium | Med Hi | Med Hi |
| Unlikely | Low | Low Med | Low Med | Medium | Med Hi |
| Very Unlikely | Low | Low | Low Med | Medium | Medium |

**Fig. 1** Probability - impact matrix. An example of use.

use appears in different sectors, ranging from the construction industry (Qazi et al. 2021), oil and gas industries (Thomas et al. 2014), to the healthcare sector (Lemmens et al. 2022), engineering projects (Koulinas et al. 2021) and, of course, project management (International Organization for Standardization, 2019; Li et al. 2018).

In a table, the risk matrix represents the probability (usually on the vertical axis of the table) and impact (usually on the horizontal axis) categories (Ale et al. 2015). These axes are further divided into different levels so that risk matrices of 3×3 levels are found with three levels set for probability and three others to define impact, 5 × 5, or even more levels (Duijm, 2015; Levine, 2012; Cox, 2008). The matrix classifies risks into different risk categories, normally labelled with qualitative indicators of severity (often colours like "Red", "Yellow" and "Green"). This classification combines each likelihood level with every impact level in the matrix (see an example of a probability-impact matrix in Fig. 1).

There are three different risk matrix typologies based on the categorisation of likelihood and impact: qualitative, semiquantitative, and quantitative. Qualitative risk matrices provide descriptive assessments of probability and consequence by establishing categories as "low," "medium" or "high" (based on the matrix's specific number of levels). In contrast, semiquantitative risk matrices represent the input categories by ascending scores, such as 1, 2, or 3 (in a 3×3 risk matrix), where higher scores indicate a stronger impact or more likelihood. Finally, in quantitative risk matrices, each category receives an assignment of numerical intervals corresponding to probability or impact estimates. For example, the "Low" probability level is associated with a probability interval [0.1 0.3] (Li et al. 2018).

Qualitative matrices classify risks according to their potential hazard, depending on where they fit into the matrix. The risk level is defined by the "colour" of the corresponding cell (in turn, this depends on the probability and impact level), with risks classified with "red" being the most important and the priority ones to pay attention to, but without distinguishing any risks in the different cells of the same colour. In contrast, quantitative risk matrices allow to classify risks according to their risk level (red, yellow, or green) and to prioritise each risk in the same colour by indicating which is the most important. Each cell is assigned a colour and a numerical value, and the product of the value is usually assigned to the probability level and the value assigned to the impact level (Risk = probability × impact).

Risk matrix use is frequent, partly due to its simple application and easy construction compared to alternative risk assessment methods (Levine, 2012). Risk matrices offer a well-defined structure for carrying out a methodical risk assessment, provide a practical justification for ranking and prioritising risks, visually and attractively inform stakeholders, among other reasons (Talbot, 2014; Ball and Watt, 2013).

However, many authors identify problems in using risk matrices (Monat and Doremus, 2020; Peace, 2017; Levine, 2012; Ni et al. 2010; Cox, 2008; Cox et al. 2005), and even the International Organization for Standardization (2019) indicates some drawbacks. The most critical problems identified in using risk matrices for strategic decision-making are that risk matrices can be inaccurate when comparing risks and they sometimes assign similar ratings to risks with significant quantitative differences. In addition, there is the risk of giving excessively high qualitative ratings to risks that are less serious from a quantitative perspective. This can lead to suboptimal decisions, especially when threats have negative correlations in frequency and severity terms. Such lack of precision can result in inefficient resource allocation because they cannot be based solely on the categories provided by risk matrices. Furthermore, the categorisation of the severity of consequences is subjective in uncertainty situations, and the assessment of probability, impact and risk ratings very much depends on subjective interpretations, which can lead to discrepancies between different users when assessing the same quantitative risks.

Given this background, several authors propose solutions to the posed problems. Goerlandt and Reniers (2016) review previous works that have attempted to respond to the problems identified with risk matrices. For example, Markowski and Mannan (2008) suggest using fuzzy sets to consider imprecision in describing ordinal linguistic scales. Subsequently, Ni et al. (2010) propose a methodology that employs probability and consequence ranks as independent score measures. Levine (2012) puts forward the use of logarithmic scales on probability and impact axes. Menge et al. (2018) recommend utilising untransformed values as scale labels due to experts' misunderstanding of logarithmic scales. Ruan et al. (2015) suggest an approach that considers decision makers' risk aversion by applying the utility theory.

Other authors, such as Duijm (2015), propose a continuous probability consequence diagram as an alternative to the risk matrix, and employing continuous scales instead of categories. They also propose utilising more comprehensive colour ranges in risk matrices whenever necessary to prioritise risks and to not simply accept them. In contrast, Monat and Doremus (2020) put forward a new risk prioritisation tool. Alternatively, Sutherland et al. (2022) suggest changing matrix size by accommodating cells' size to the risk's importance. Even Proto et al. (2023) recommend avoiding colour in risk matrices so that the provided information is unbiased due to the bias that arises when using coloured matrices.

By bearing in mind the difficulties presented by the results offered by risk matrices, we propose a quantitative method for risk prioritisation. We use qualitative risk analysis data by maintaining the estimate of the probability of each risk occurring and its potential impact. Nevertheless, instead of entering these data into the risk matrix, our project model contains them for Monte Carlo simulation. As a result, we obtain a quantified prioritisation of each risk that differentiates the importance of each risk according to the impact on cost and duration objectives.

## Methodology

Figure 2 depicts the proposed method for prioritising project risks using quantitative techniques. At the end of the process, and with the prioritised risks indicating the absolute value of the impact of each risk on the project, the organisation can efficiently allocate resources to the risks identified as the most critical ones.

The top of the diagram indicates the risk phases that belong to the overall risk management process. Below them it reflects the steps of the proposed model that would apply in each phase.

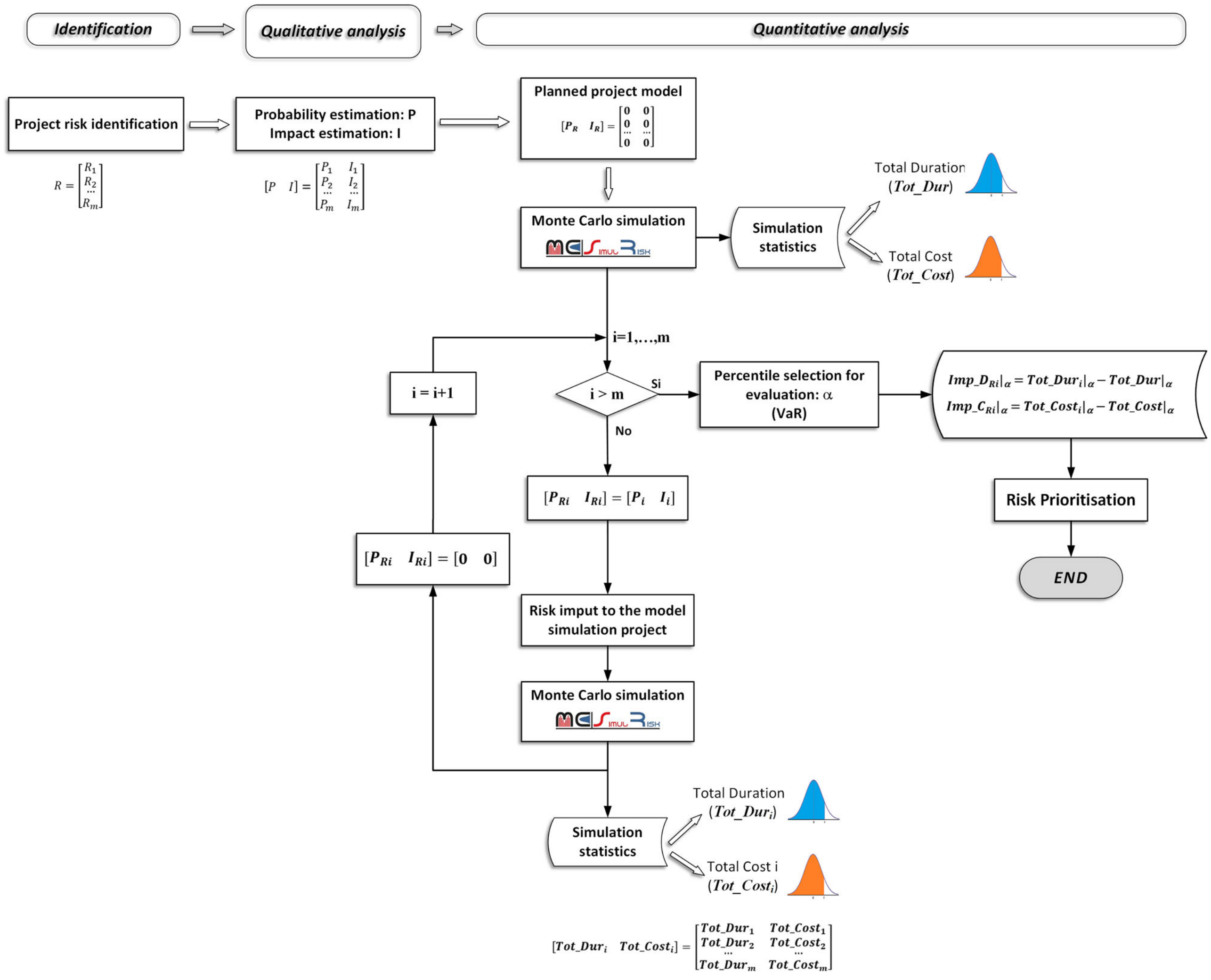

**Fig. 2** Quantitative Risk Assessment Flow Chart.

The first step corresponds to the project's "**risk identification**". Using the techniques or tools established by the organisation (brainstorming, Delphi techniques, interviews, or others), we obtain a list of the risks ($R$) that could impact the project objectives (Eq. 1), where $m$ is the number of risks identified in the project.

$$R = \begin{bmatrix} R_1 \\ R_2 \\ \dots \\ R_m \end{bmatrix} \tag{1}$$

Next we move on to the "**risk estimation**" phase, in which a distribution function must be assigned to the probability that each identified risk will appear. We also assign the distribution function associated with the risk's impact. Traditionally, the qualitative risk analysis defines semantic values (low, medium, high) to assign a level of probability and risk impact. These semantic values are used to evaluate the risk in the probability-impact matrix. Numerical scales apply in some cases, which help to assign a semantic level to a given risk (Fig. 3).

Our proposed model includes the three uncertainty types put forward by Hillson (2014), namely aleatoric, stochastic and epistemic, to identify and assess different risks. Ontological uncertainty is not considered because it goes beyond the limits of human knowledge and cannot, therefore, be modelled (Alleman et al. 2018a).

| SCALE | PROBABILITY | +/- IMPACT ON PROJECT OBJECTIVES | | |
|---|---|---|---|---|
| | | TIME | COST | QUALITY |
| Very High | >70% | >6 months | >$5M | Very significant impact on overall functionality |
| High | 51-70% | 3-6 months | $1M-$5M | Significant impact on overall functionality |
| Medium | 31-50% | 1-3 months | $501k-$1M | Some impact in key functional areas |
| Low | 11-30% | 1-4 weeks | $100k-$500k | Minor impact on overall functionality |
| Very Low | 1-10% | 1 week | <$100k | Minor impact on secondary functions |
| Nil | <1% | No change | No change | No change in functionality |

**Fig. 3 Correspondence of the semantic level to risk probability and impact.** Source: Project Management Institute (2017).

A risk can have aleatoric uncertainty as regards the probability of its occurrence, and mainly for its impact if its value can fluctuate over a set range due to its variability. This aleatoric risk uncertainty can be modelled using a probability distribution function (PDF), exactly as we do when modelling activity uncertainty (Acebes et al. 2015, 2014). As the risk management team's (or project management team's) knowledge of the project increases, and as more information about the risk becomes available, the choice of the PDF (normal, triangular, beta, among others) and its parameters become more accurate.

A standard definition of risk is "an uncertain event that, if it occurs, may impact project objectives" (Project Management

Institute, 2017). A risk, if defined according to the above statement, perfectly matches the stochastic uncertainty definition proposed by Hillson (2014). Moreover, one PDF that adequately models this type of uncertainty is a Bernoulli distribution function (Vose, 2008). Thus for deterministic risk probability estimates (the same as for risk impact), we model this risk (probability and impact) with a Bernoulli-type PDF that allows us to introduce this type of uncertainty into our simulation model.

Finally, epistemic uncertainties remain to be modelled, such as those for which we do not have absolute information about and that arise from a lack of knowledge (Damnjanovic and Reinschmidt, 2020; Alleman et al. 2018b). In this case, risks (in likelihood and impact terms) are classified into different levels, and all these levels are assigned a numerical scale (as opposed to the methodology used in a qualitative risk analysis, where levels are classified with semantic values: "high", "medium" and "low").

"Epistemic uncertainty is characterised by not precisely knowing the probability of occurrence or the magnitude of a potential impact. Traditionally, this type of risk has been identified with a qualitative term: "Very Low", "Low", "Medium", "High" and "Very High" before using the probability-impact matrix. Each semantic category has been previously defined numerically by identifying every numerical range with a specific semantic value (Bae et al. 2004). For each established range, project managers usually know the limits (upper and lower) between which the risk (probability or impact) can occur. However, they do not certainly know the value it will take, not even the most probable value within that range. Therefore, we employ a uniform probability function to model epistemic uncertainty (i.e., by assuming that the probability of risk occurrence lies within an equiprobable range of values). Probabilistic representations of uncertainty have been successfully employed with uniform distributions to characterise uncertainty when knowledge is sparse or absent (Curto et al. 2022; Vanhoucke, 2018; Helton et al. 2006).

The choice of the number and range of each level should be subject to a thorough analysis and consideration by the risk management team. As each project is unique, there are ranges within which this type of uncertainty can be categorised. Different ranges apply to assess likelihood and impact. Furthermore for impact, further subdivision helps to distinguish between impact on project duration and impact on project costs. For example, when modelling probability, we can set five probability levels corresponding to intervals: [0 0.05], [0.05 0.2], [0.2 0.5], and so on. With the time impact, for example, on project duration, five levels as follows may apply: [0 1], [1 4], [4 12], …. (measured in weeks, for example).

Modelling this type of uncertainty requires the risk management team's experience, the data stored on previous projects, and constant consultation with project stakeholders. The more project knowledge available, the more accurate the proposed model is for each uncertainty, regardless of it lying in the number of intervals, their magnitude or the type of probability function (PDF) chosen to model that risk.

Some authors propose using uniform distribution functions to model this type of epistemic uncertainty because it perfectly reflects lack of knowledge about the expected outcome (Eldosouky et al. 2014; Vose, 2008). On the contrary, others apply triangular functions, which require more risk knowledge (Hulett, 2012). Following the work by Curto et al. (2022), we employ uniform distribution functions.

As a result of this phase, we obtain the model and the parameters that model the distribution functions of the probability ($P$) and impact ($I$) of each identified risk in the previous phase (Eq. 2).

$$\begin{bmatrix} P & I \end{bmatrix} = \begin{bmatrix} P_1 & I_1 \\ P_2 & I_2 \\ \cdots & \cdots \\ P_m & I_m \end{bmatrix} \tag{2}$$

Once the risks identified in the project have been defined and their probabilities and impacts modelled, we move on to "**quantitative risk prioritisation**". We start by performing MCS on the planned project model by considering only the aleatoric uncertainty of activities. In this way, we learn the project's total duration and cost, which is commonly done in a Monte Carlo analysis. In Monte Carlo Methods (MCS), expert judgement and numerical methods are combined to generate a probabilistic result through simulation routine (Ammar et al. 2023). This mathematical approach is noted for its ability to analyse uncertain scenarios from a probabilistic perspective. MCS have been recognised as outperforming other methods due to their accessibility, ease of use and simplicity. MCS also allow the analysis of opportunities, uncertainties, and threats (Al-Duais and Al-Sharpi, 2023). This technique can be invaluable to risk managers and helpful for estimating project durations and costs (Ali Elfarra and Kaya, 2021).

As inputs to the simulation process, we include defining project activities (duration, cost, precedence relationship). We also consider the risks identified in the project, which are those we wish to prioritise and to obtain a list ordered by importance (according to their impact on not only duration, but also on project cost). The 'MCSimulRisk' software application (Acebes, Curto, et al. 2023; Acebes, De Antón, et al. 2023) allows us to perform MCS and to obtain the main statistics that result from simulation (including percentiles) that correspond to the total project duration ($Tot_Dur$) and to its total cost ($Tot_Cost$) (Eq. 3).

$$\begin{bmatrix} Tot_Dur & Tot_Cost \end{bmatrix} \tag{3}$$

Next, we perform a new simulation by including the first of the identified risks ($R_1$) in the project model, for which we know its probability ($P_1$) and its Impact ($I_1$). After MCS, we obtain the statistics corresponding to this simulation ([$Tot_Dur_1$ $Tot_Cost_1$]). We repeat the same operation with each identified risk ($R_i$, $i = 1, \ldots, m$) and obtain the main statistics corresponding to each simulation (Eq. 4).

$$\begin{bmatrix} Tot_Dur_i & Tot_Cost_i \end{bmatrix} = \begin{bmatrix} Tot_Dur_1 & Tot_Cost_1 \\ Tot_Dur_2 & Tot_Cost_2 \\ \cdots & \cdots \\ Tot_Dur_m & Tot_Cost_m \end{bmatrix} \tag{4}$$

Once all simulations (the same number as risks) have been performed, we must choose a confidence percentile to calculate risk prioritisation (Rezaei et al. 2020; Sarykalin et al. 2008). Given that the total duration and cost results available to us, obtained by MCS, are stochastic and have variability (they are no longer constant or deterministic), we must choose a percentile (α) that conveys the risk appetite that we are willing to assume when calculating. Risk appetite is "*the amount and type of risk that an organisation is prepared to pursue, retain or take*" (International Organization for Standardization, 2018).

A frequently employed metric for assessing risk in finance is the Value at Risk (VaR) (Caron, 2013; Caron et al. 2007). In financial terms, it is traditional to choose a P95 percentile as risk appetite (Chen and Peng, 2018; Joukar and Nahmens, 2016; Gatti et al. 2007; Kuester et al. 2006; Giot and Laurent, 2003). However in project management, the P80 percentile is sometimes chosen as the most appropriate percentile to measure risk appetite (Kwon

**Table 2 Definition of project activities (duration and cost) from a real-life project.**

| Activity | Predecessor | Duration (days) | | | Cost (x 1,000 monetary units) | |
|---|---|---|---|---|---|---|
| | | Min | Mp | Max | FC | VC |
| **Ai** | | - | - | - | - | - |
| **Engineering** | | | | | | |
| **A1** | Ai | 10 | 15 | 20 | 0 | 1.0 |
| **A2** | A1 | 25 | 30 | 35 | 0 | 2.5 |
| **A3** | A1 | 20 | 30 | 40 | 0 | 4.5 |
| **A4** | A2, A3 | 45 | 55 | 65 | 0 | 4.5 |
| **A5** | A2 | 60 | 70 | 80 | 0 | 7.5 |
| **A6** | A2 | 80 | 90 | 100 | 0 | 3.5 |
| **A7** | A1 | 50 | 70 | 90 | 0 | 10 |
| **Procurement** | | | | | | |
| **A8** | A5 | 20 | 25 | 30 | 650.5 | 0 |
| **A9** | A5 | 70 | 85 | 100 | 4200 | 0 |
| **A10** | A6 | 70 | 85 | 100 | 3675 | 0 |
| **A11** | A7 | 70 | 80 | 90 | 7000 | 0 |
| **A12** | A7 | 100 | 110 | 120 | 75 | 0 |
| **A13** | A7 | 70 | 80 | 90 | 500 | 14.4 |
| **A14** | A11, A13 | 25 | 30 | 35 | 100 | 6 |
| **A15** | A14 | 12 | 15 | 18 | 0 | 6 |
| **A16** | A1 | 170 | 190 | 210 | 300 | 1.8 |
| **Construction** | | | | | | |
| **A17** | A2 | 45 | 55 | 65 | 300 | 4.8 |
| **A18** | A17 | 50 | 60 | 70 | 550 | 5.4 |
| **A19** | A3 | 45 | 55 | 65 | 575 | 6 |
| **A20** | A3, A4 | 80 | 95 | 110 | 675 | 5.4 |
| **A21** | A18 | 20 | 25 | 30 | 275 | 3.6 |
| **A22** | A21 | 18 | 20 | 22 | 275 | 3.6 |
| **A23** | A18 | 35 | 40 | 45 | 287 | 4.8 |
| **A24** | A21 | 30 | 40 | 50 | 275 | 3 |
| **A25** | A8 | 35 | 45 | 55 | 75 | 8 |
| **A26** | A21, A25 | 75 | 85 | 95 | 125 | 9.6 |
| **A27** | A9, A22, A23 | 75 | 80 | 85 | 100 | 8 |
| **A28** | A10, A24 | 45 | 60 | 75 | 375 | 25 |
| **A29** | A12, A20 | 12 | 15 | 18 | 0 | 2 |
| **A30** | A15, A16, A29 | 50 | 60 | 70 | 75 | 7 |
| **A31** | A26, A30 | 12 | 15 | 18 | 0 | 3 |
| **A32** | A27, A28, A31 | 12 | 15 | 18 | 0 | 3 |
| **Af** | A32 | - | - | - | - | - |

and Kang, 2019; Traynor and Mahmoodian, 2019; Lorance and Wendling, 2001).

Finally, after choosing the risk level we are willing to assume, we need to calculate how each risk impacts project duration ($Imp_D_{Ri}$) and costs ($Imp_C_{Ri}$). To do so, we subtract the original value of the total project expected duration and costs (excluding all risks) from the total duration and costs of the simulation in which we include the risk we wish to quantify (Eq. 5).

$$\begin{aligned} Imp_D_{Ri}\big|_a &= Tot_Dur_i\big|_a - Tot_Dur\big|_a \\ Imp_C_{Ri}\big|_a &= Tot_Cost_i\big|_a - Tot_Cost\big|_a \end{aligned} \quad (5)$$

Finally, we present these results on two separate lists, one for the cost impact and one for the duration impact, by ranking them according to their magnitude.

## Case study

In this section, we use a real-life project to illustrate how to apply the proposed method for quantitative risk prioritisation purposes. For this purpose, we choose an engineering, procurement and construction project undertaken in South America and used in the literature by Votto et al. (2020a, 2020b).

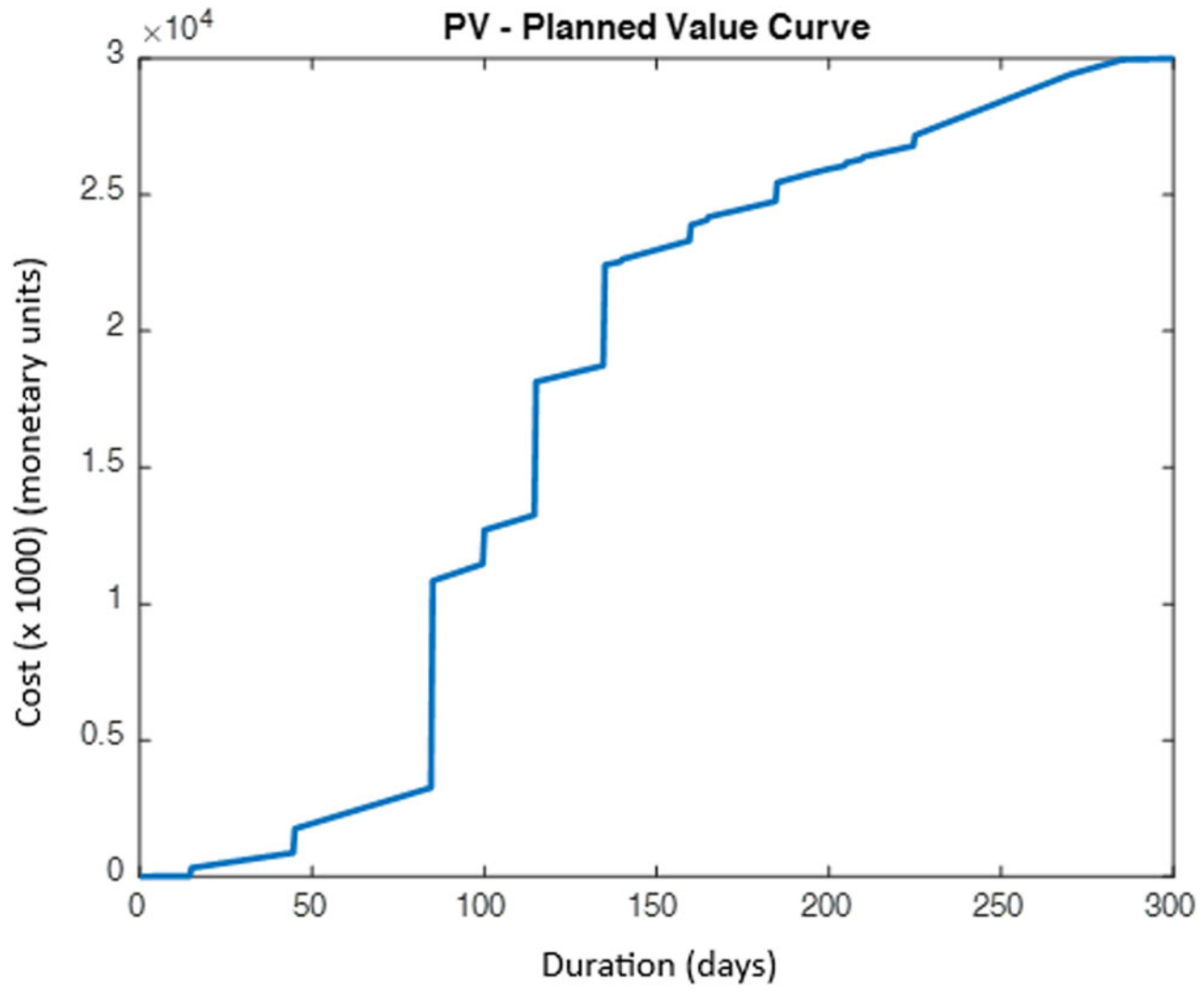

**Fig. 4** Planned value curve of the real-life project.

**Project description**. The project used as an application example consists of the expansion of an industrial facility. It covers a wide spectrum of tasks, such as design and engineering work, procurement of machinery and its components, civil construction, installation of all machinery, as well as commissioning and starting up machines (Votto et al. 2020a, 2020b).

Table 2 details the parameters that we use to define activities. The project comprises 32 activities, divided into three groups: engineering, procurement and construction (EPC). A fictitious initial activity (*Ai*) and a fictitious final activity (*Af*) are included. We employ triangular distribution functions, whose parameters are the minimum value (*Min*), the most probable value (*Mp*) and the maximum value (*Max*), to model the random duration of activities, expressed as days. We divide the cost of each activity in monetary units into a fixed cost (*FC*), independently of activity duration, and the variable cost (*VC*), which is directly proportional to project duration. As activity duration can vary, and the activity cost increases directly with its duration, the total project cost also exhibits random variations.

Under these conditions, the planned project duration is 300 days and has a planned cost of 30,000 (x1000) monetary units. Figure 4 shows the Planned Value Curve of the project.

The next step in the methodology (Fig. 2) is to identify the project risks. To do this, the experts' panel meets, analyses all the project documentation. Based on their personal experience with other similar projects and after consulting all the involved stakeholders, it provides a list of risks (see Table 3).

It identifies 11 risks, of which nine have the potential to directly impact the project duration objective (R1 to R9), while six may impact the cost objective (R10 to R15). The risks that might impact project duration and cost have two assigned codes. We identify the project phase and activity on which all the identified risks may have an impact (Table 3).

The next step is to estimate the likelihood and impact of the identified risks (qualitative analysis). Having analysed the project and consulted the involved stakeholders, the team determines the project's different probability and impact levels (duration and cost). The estimation of these ranges depends on the project budget, the estimated project duration, and the team's experience in assigning the different numerical values to each range. As a result, the project team is able to construct the probability-impact matrix shown in Fig. 5.

Each probability range for risk occurrence in this project is defined. Thus for a very low probability (VL), the assigned probability range is between 0 and 3% probability, for a low level (L), the assigned range lies between 3% and 10% probability of risk occurrence, and so on with the other established probability ranges (medium, high, very high).

**Table 3 Identification of the risks from the real-life project.**

| Risk | Risk Id. | | Phase | Prob. | Impact (I) | | Act. Imp |
|---|---|---|---|---|---|---|---|
| | Dur | Cost | | (P) | Dur | Cost | |
| Technical complexity of the engineering project | R1 | | Engineering | VL | L | - | 6 |
| Change of requirements in the project | R2 | R10 | Engineering | H | M | L | 2 |
| Inadequate supplier management | | R11 | Procurement | L | - | M | 8 |
| Supply chain disruptions | R3 | R12 | Procurement | L | H | L | 13 |
| Inaccurate internal needs analysis | R4 | | Procurement | M | L | - | 12 |
| Health and safety hazards | R5 | | Construction | M | M | - | 20 |
| Adverse weather conditions | R6 | R13 | Construction | L | L | M | 26 |
| Difficulties in obtaining permits and licences | R7 | R14 | Construction | VL | M | VL | 30 |
| Availability of skilled labour | R8 | | Construction | M | M | - | 28 |
| Availability of construction materials on time | R9 | | Construction | L | M | - | 23 |
| Changes in the price of raw materials | | R15 | Construction | L | - | H | 31 |

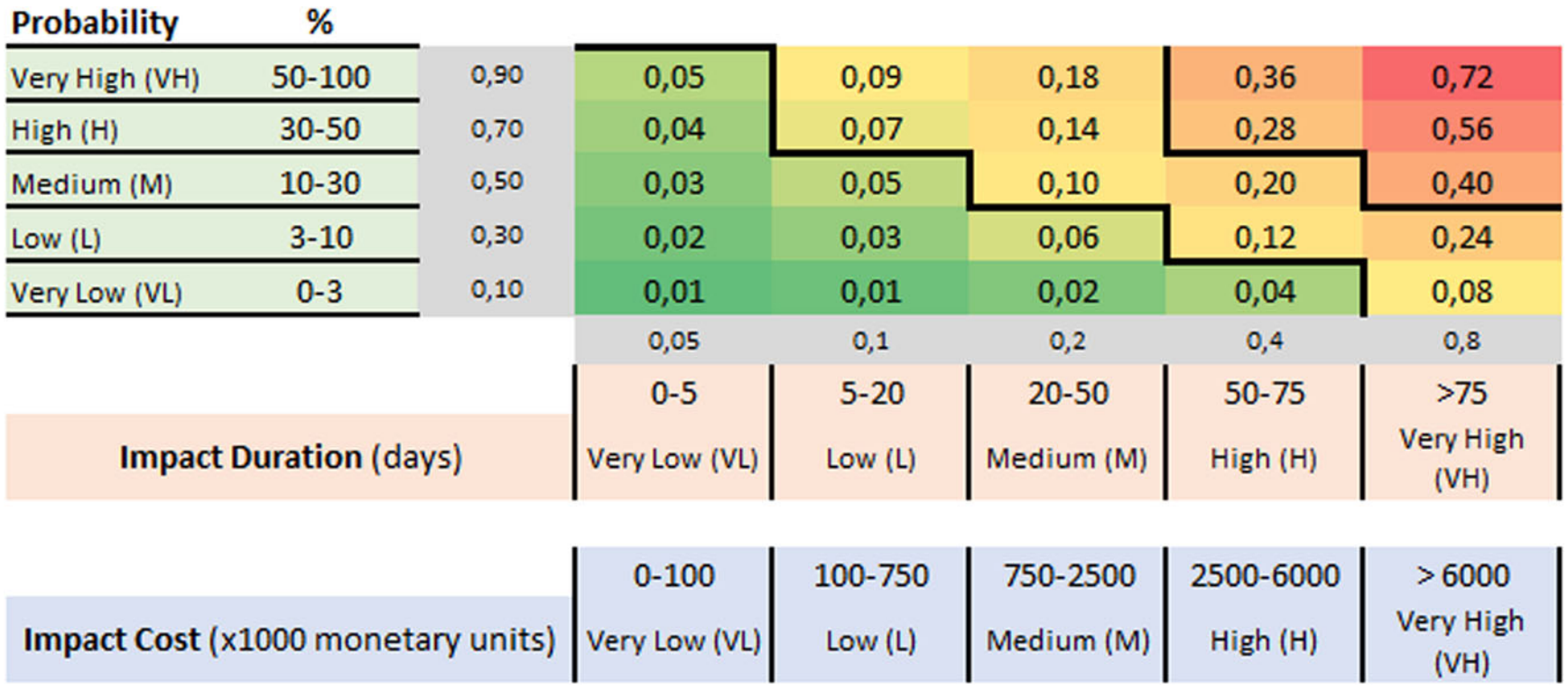

**Fig. 5** Estimation of the probability and impact ranges.

The different impact ranges are also defined by differentiating between impacts in duration and cost terms. Thus a VL duration impact is between 0 and 5 days, while the same range (VL) in cost is between 0 and 100 (x1000) monetary units. Figure 5 shows the other ranges and their quantification in duration and cost terms.

The combination of each probability level and every impact level coincides in a cell of the risk matrix (Fig. 5) to indicate the risk level ("high", "medium", and "low") according to the qualitative analysis. Each cell is assigned a numerical value by prioritising the risks at the same risk level. This work uses the matrix to compare the risk prioritisation results provided by this matrix to those provided by the proposed quantitative method.

A probability and impact value are assigned to each previously identified risk (Table 3). Thus, for example, for the risk called "Interruptions in the supply chain", coded as R3 for impacting activity 13 duration, we estimate an L probability and a strong impact on duration (H). As this same risk might impact the activity 13 cost, it is also coded as R12, and its impact on cost is estimated as L (the probability is the same as in R3; Table 3).

Finally, to conclude the proposed methodology and to prioritise the identified risks, we use the "MCSimulRisk" software application by incorporating MCS (in this work, we employ 20,000 iterations in each simulation). Activities are modelled using triangular distribution functions to incorporate project information into the simulation application. Costs are modelled with fixed and variable costs depending on the duration of the corresponding activity. Furthermore, risks (probability and impact) are modelled by uniform distribution functions. Figure 6 depicts the project network and includes the identified risks that impact the corresponding activities.

## Results and discussion

In order to obtain the results of prioritising the identified risks, we must specify a percentile that determines our risk aversion. This is the measure by which we quantify the risk. Figure 7 graphically justifies the choice of P95 as a risk measure, as opposed to a lower percentile, which corroborates the view in the literature and appears in **Methodology**. In Fig. 7, we plot the probability distribution and cumulative distribution functions corresponding to the total project planned cost, together with the cost impact of one of the risks. The impact caused by the risk on the total cost corresponds to the set of iterations whose total cost is higher than that planned (bottom right of the histogram).

By choosing P95 as VaR, we can consider the impact of a risk on the project in the measure. In this example, for P95 we obtain a total cost value of $3.12 \times 10^7$ monetary units. Choosing a lower percentile, e.g. P80, means that the value we can obtain with this choice can be considerably lower ($3.03 \times 10^7$ monetary units), and might completely ignore the impact of the risk on the total project cost. However, project managers can choose the percentile that represents their risk aversion.

Once the percentile on which to quantify the risk is chosen, the "MCSimulRisk" application provides us with the desired results for prioritising project risks (Fig. 8). For the chosen percentile (P95), which represents our risk appetite for this project, the planned project duration is 323.43 days. In other words, with a 95% probability the planned project will be completed before 323.43 days. Similarly, the P95 corresponding to cost is 30,339 ×1000 monetary units. The application also provides us with the project duration in the first column of Fig. 8 after incorporating all the identified risks (corresponding to a P95 risk appetite) into the planned project. Column 2 of the same figure

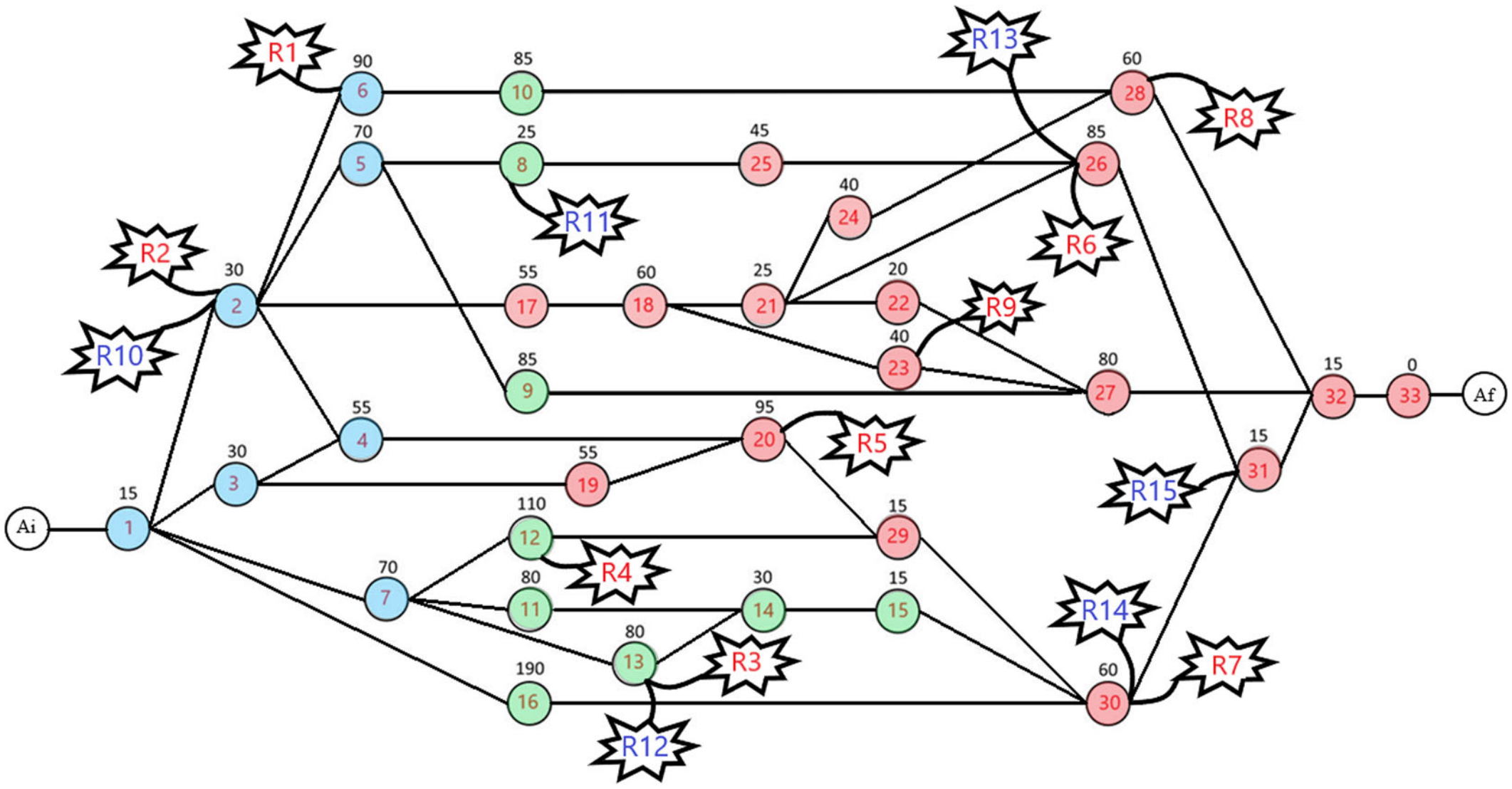

**Fig. 6** Network diagram of the project together with the identified risks.

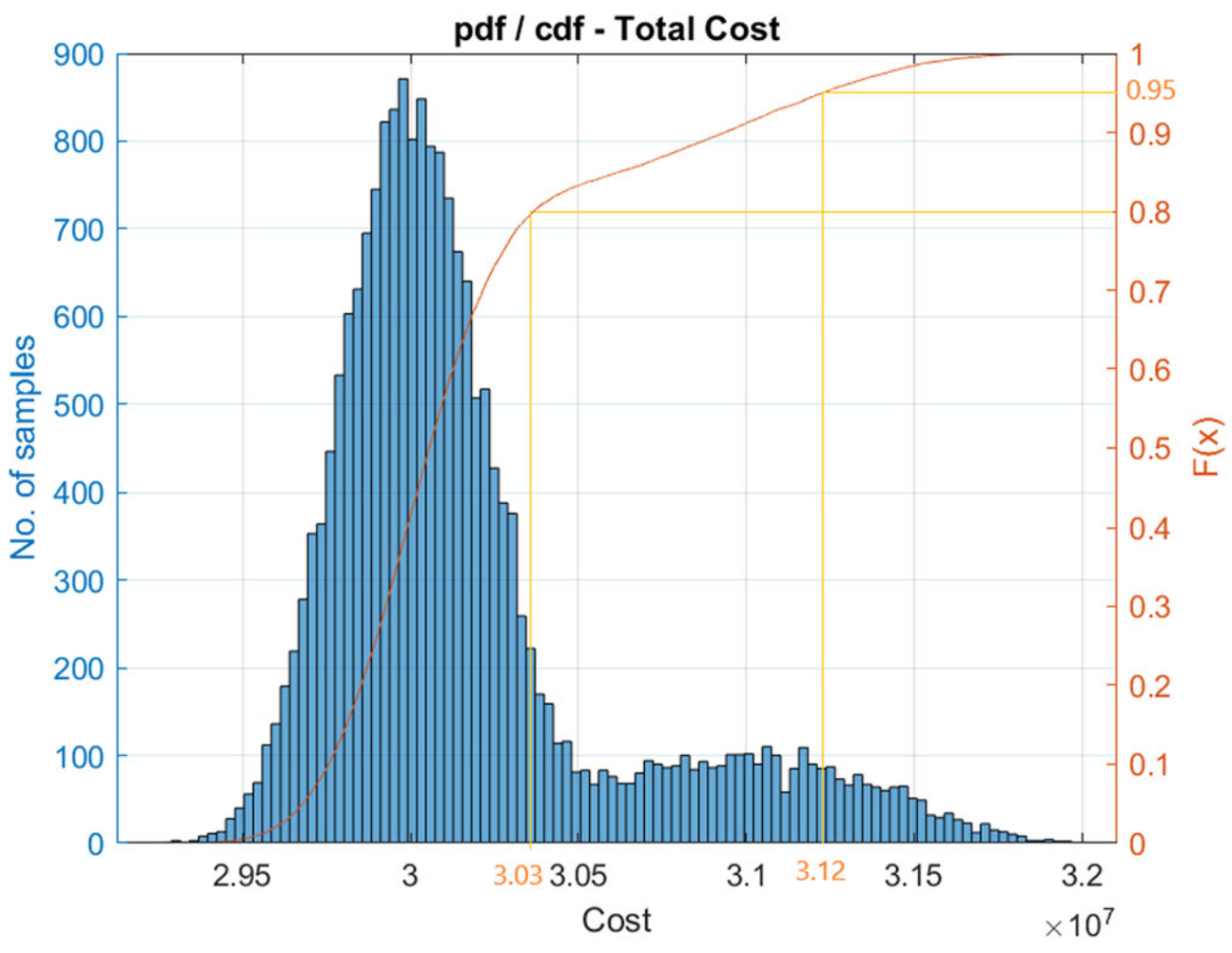

**Fig. 7 Distribution function and cumulative distribution of the Total Planned Cost and cost risk impact.** Source: MCSimulRisk.

shows the project cost after incorporating the corresponding risk into the model.

With the results in the first two columns (total project duration and cost after incorporating the corresponding risks), and by knowing the planned total project duration and cost (without considering risks) for a given percentile (P95), we calculate the values of the following columns in Fig. 8. Thus column 3 represents the difference between the planned total project duration value (risk-free) and project duration by incorporating the corresponding risk that we wish to quantify. Column 4 prioritises the duration risks by ranking according to the duration that each risk contributes to the project. Column 5 represents the difference between the planned total project cost (risk-free) and the total project cost by incorporating the corresponding risk. Finally, Column 6 represents the ranking or prioritisation of the project risks according to their impact on cost.

To compare the results provided by this methodology in this paper we propose quantitative risk prioritisation, based on MCS. We draw up Table 4 with the results provided by the probability-impact matrix (Fig. 5).

The first set of columns in Table 4 corresponds to the implementation of the risk matrix (probability-impact matrix) for the identified risks. The second group of columns represents the prioritisation of risks according to their impact on duration (data obtained from Fig. 8). The third group corresponds to the risk prioritisation according to their impact on cost (data obtained from Fig. 8).

For the project proposed as an example, we find that risk R3 is the most important one if we wish to control the total duration because it corresponds to the risk that contributes the most duration to the project if it exists. We note that risks R10 to R15 do not impact project duration. If these risks materialise, their contribution to increase (or decrease, as the case may be) project duration is nil.

On the impact on project costs, we note that risk R15 is the most important. It is noteworthy that risk R5 is the fourth most important risk in terms of impact on the total project costs, even though it is initially identified as a risk that impacts project duration. Unlike cost risks (which do not impact the total project duration), the risks that can impact project duration also impact total costs.

We can see that the order of importance of the identified risks differs depending on our chosen method (risk matrix *versus* quantitative prioritisation). We independently quantify each risk's impact on the cost and duration objectives. We know not only the order of importance of risks (R3, R5, etc.) but also the magnitude of their impact on the project (which is the absolute delay caused by a risk in duration terms or what is the absolute cost overrun generated by a risk in cost terms). It seems clear that one risk is more important than another, not only because of the estimation of its probability and impact but also because the activity on which it impacts may have a high criticality index or not (probability of belonging to the project's critical path).

As expected, the contribution to the total duration of the identified risks that impact only cost is zero. The same is not valid for the risks identified to have an impact on duration because the latter also impacts the cost objective. We also see how the risks that initially impact a duration objective are more critical for their impact on cost than others that directly impact the project's cost (e.g. R5).

## Conclusions

The probability-impact matrix is used in project management to identify the risk to which the most attention should be paid

```
The Project Duration for this percentile is   323.4339

The Project Cost for this percentile is   3.0339e+04


Quantitative_Prioritisation_of_Risks =

  15×6 table
```

| | Duration_with_Ri | Cost_with_Ri | Difference_Duration_with_Ri | Ranking_Dur | Difference_Cost_with_Ri | Ranking_Cost |
|---|---|---|---|---|---|---|
| R1 | 323.77 | 30340 | 0.33481 | 9 | 0.2357 | 13 |
| R2 | 346.07 | 30408 | 22.639 | 3 | 68.359 | 7 |
| R3 | 349.76 | 30339 | 26.33 | 1 | 0 | 14 |
| R4 | 325.49 | 30339 | 2.0554 | 5 | 0 | 14 |
| R5 | 346.1 | 31090 | 22.666 | 2 | 750.36 | 4 |
| R6 | 324.19 | 30352 | 0.76083 | 8 | 13.16 | 9 |
| R7 | 324.45 | 30349 | 1.0162 | 6 | 9.6971 | 10 |
| R8 | 338.25 | 30396 | 14.818 | 4 | 56.726 | 8 |
| R9 | 324.3 | 30346 | 0.86284 | 7 | 6.2513 | 11 |
| R10 | 323.43 | 30753 | 0 | 10 | 413.55 | 5 |
| R11 | 323.43 | 31106 | 0 | 10 | 766.54 | 3 |
| R12 | 323.43 | 30421 | 0 | 10 | 81.405 | 6 |
| R13 | 323.43 | 31117 | 0 | 10 | 777.35 | 2 |
| R14 | 323.43 | 30340 | 0 | 10 | 0.72532 | 12 |
| R15 | 323.43 | 33400 | 0 | 10 | 3060.9 | 1 |

**Fig. 8 Monte Carlo simulation results with "MCSimulRisk".** The first column corresponds to the risks identified. Columns *Duration_with_Ri* and *Cost_with_Ri* represent the simulation values, including the corresponding risk. Columns *Difference_Duration_with_Ri and Difference_Cost_with_Ri* represent the difference in duration and cost of each simulation concerning the value obtained for the chosen percentile. Finally, *Ranking_Dur* and *Ranking_Cost* represent the prioritisation of risks in duration and cost, respectively.

**Table 4 Comparison of risk prioritisation: Probability-Impact Matrix & Proposed Method.**

| | Probability – Impact Matrix | | | | Duration (days) | | | Cost x 1,000 (monetary units) | | |
|---|---|---|---|---|---|---|---|---|---|---|
| Id. Risk | P | I | P x I | Ranking | Dur_with_R | Diff_Dur | Ranking_Dur | Cost_with_R | Diff_Cost | Ranking_Cost |
| R1 | MB | B | 0.01 | 9 | 323.77 | 0.335 | 9 | 30,339.55 | 0.236 | 13 |
| R2 | A | M | 0.14 | 1 | 346.07 | 22.639 | 3 | 30,407.68 | 68.359 | 7 |
| R3 | B | A | 0.12 | 2 | 349.76 | 26.330 | 1 | 30,339.32 | 0.000 | 14 |
| R4 | M | B | 0.05 | 6 | 325.49 | 2.055 | 5 | 30,339.32 | 0.000 | 14 |
| R5 | M | M | 0.1 | 3 | 346.10 | 22.666 | 2 | 31,089.68 | 750.362 | 4 |
| R6 | B | B | 0.03 | 7 | 324.19 | 0.761 | 8 | 30,352.48 | 13.160 | 9 |
| R7 | MB | M | 0.02 | 8 | 324.45 | 1.016 | 6 | 30,349.01 | 9.697 | 10 |
| R8 | M | M | 0.1 | 3 | 338.25 | 14.818 | 4 | 30,396.04 | 56.726 | 8 |
| R9 | B | M | 0.06 | 5 | 324.30 | 0.863 | 7 | 30,345.57 | 6.251 | 11 |
| R10 | A | B | 0.07 | 2 | 323.43 | 0 | 10 | 30,752.87 | 413.553 | 5 |
| R11 | B | M | 0.06 | 3 | 323.43 | 0 | 10 | 31,105.85 | 766.538 | 2 |
| R12 | B | B | 0.03 | 5 | 323.43 | 0 | 10 | 30,420.72 | 81.405 | 6 |
| R13 | B | M | 0.06 | 3 | 323.43 | 0 | 10 | 31,116.67 | 777.350 | 3 |
| R14 | MB | MB | 0.01 | 6 | 323.43 | 0 | 10 | 30,340.04 | 0.725 | 12 |
| R15 | B | A | 0.12 | 1 | 323.43 | 0 | 10 | 33,400.24 | 3,060.927 | 1 |

during project execution. This paper studies how the risk matrix is adopted by a large majority of standards, norms and methodologies in project management and, at the same time, practitioners and academics recognise it as a fundamental tool in the qualitative risks analysis.

However, we also study how this risk matrix presents particular problems and offers erroneous and contradictory results. Some studies suggest alternatives to its use. Notwithstanding, it continues to be a widely employed tool in the literature by practitioners and academics. Along these lines, with this work we propose an alternative to the probability-impact matrix as a tool to know the most critical risk for a project that can prevent objectives from being fulfilled.

For this purpose, we propose a quantitative method based on MCS, which provides us with numerical results of the importance of risks and their impact on total duration and cost objectives. This proposed methodology offers significant advantages over other risk prioritisation methods and tools, especially the traditional risk matrix. The proposed case study reveals that risk prioritisation yields remarkably different results depending on the selected method, as our findings confirm.

In our case, we obtain numerical values for the impact of risks on total duration and cost objectives, and independently of one another. This result is interesting for project managers because they can focus decision-making on the priority order of risks and the dominant project objective (total duration or total cost) if they do not coincide.

From the obtained results, we find that the risks with an impact on the cost of activities do not influence the total duration result. The risks that impact project duration also impact the total cost target. This impact is more significant than that of a risk that impacts only the activity's cost. This analysis leads us to believe that this quantitative prioritisation method has incredible potential for academics to extend their research on project risks and for practitioners to use it in the day-to-day implementation of their projects.

The proposed methodology will allow project managers to discover the most relevant project risks so they can focus their control efforts on managing those risks. Usually, implementing risk response strategies might be expensive (control efforts, insurance contracts, preventive actions, or others). Therefore, it is relevant to concentrate only on the most relevant risks. The proposed methodology allows project managers to select the most critical risks by overcoming the problems exhibited by previous methodologies like the probability-impact matrix.

In addition to the above, the risk prioritisation achieved by applying the proposed methodology is based on quantifying the impacts that risks may have on the duration and cost objectives of the project. Finally, we achieve an independent risk prioritisation in duration impact and project cost impact terms. This is important because the project manager can attach more importance to one risk or other risks depending on the priority objective that predominates in the project, the schedule or the total cost.

Undoubtedly, the reliability of the proposed method depends mainly on the accuracy of estimates, which starts by identifying risks and ends with modelling the probability and impact of each risk. The methodology we propose in this paper overcomes many of the problems of previous methodologies, but still has some limitations for future research to deal with. First of all, the results of simulations depend on the estimations of variables (probability distributions and their parameters, risk aversion parameters, etc.). Methodologies for improving estimations are beyond the scope of this research; we assume project teams are sufficient experts to make rational estimationsbased on experience and previous knowledge. Secondly, as risks are assumed to be independent, the contribution or effect of a particular risk can be estimated by including it in simulation and by computing its impact on project cost and duration. This is a reasonable assumption for most projects. In some very complex projects, however, risks can be related to one another. Further research should be done to face this situation.

As an additional research line, we plan to conduct a sensitivity study by simulating many different projects to analyse the robustness of the proposed method.

Finally, it is desirable to implement this methodology in real projects and see how it responds to the reality of a project in, for example, construction, industry, or any other sector that requires a precise and differentiated risk prioritisation.

## Data availability

Data will be made available on request.

## References

Acebes F, Curto D, De Antón J, Villafáñez, F (2024) Análisis cuantitativo de riesgos utilizando "MCSimulRisk" como herramienta didáctica. *Dirección y Organización*, *82(Abril 2024)*, 87–99. https://doi.org/10.37610/dyo.v0i82.662

Acebes F, De Antón J, Villafáñez F, Poza, D (2023) A Matlab-Based Educational Tool for Quantitative Risk Analysis. In *IoT and Data Science in Engineering Management* (Vol. 160). Springer International Publishing. https://doi.org/10.1007/978-3-031-27915-7_8

Acebes F, Pajares J, Galán JM, López-Paredes A (2014) A new approach for project control under uncertainty. Going back to the basics. Int J Proj Manag 32(3):423–434. https://doi.org/10.1016/j.ijproman.2013.08.003

Acebes F, Pereda M, Poza D, Pajares J, Galán JM (2015) Stochastic earned value analysis using Monte Carlo simulation and statistical learning techniques. Int J Proj Manag 33(7):1597–1609. https://doi.org/10.1016/j.ijproman.2015.06.012

Al-Duais FS, Al-Sharpi RS (2023) A unique Markov chain Monte Carlo method for forecasting wind power utilizing time series model. Alex Eng J 74:51–63. https://doi.org/10.1016/j.aej.2023.05.019

Ale B, Burnap P, Slater D (2015) On the origin of PCDS - (Probability consequence diagrams). Saf Sci 72:229–239. https://doi.org/10.1016/j.ssci.2014.09.003

Ali Elfarra M, Kaya M (2021) Estimation of electricity cost of wind energy using Monte Carlo simulations based on nonparametric and parametric probability density functions. Alex Eng J 60(4):3631–3640. https://doi.org/10.1016/j.aej.2021.02.027

Alleman GB, Coonce TJ, Price RA (2018a) Increasing the probability of program succes with continuous risk management. Coll Perform Manag, Meas N. 4:27–46

Alleman GB, Coonce TJ, Price RA (2018b) What is Risk? Meas N. 01(1):25–34

Ammar T, Abdel-Monem M, El-Dash K (2023) Appropriate budget contingency determination for construction projects: State-of-the-art. Alex Eng J 78:88–103. https://doi.org/10.1016/j.aej.2023.07.035

Axelos (2023) *Managing Successful Projects with PRINCE2® 7th ed.* (AXELOS Limited, Ed.; 7th Ed). TSO (The Stationery Office)

Bae HR, Grandhi RV, Canfield RA (2004) Epistemic uncertainty quantification techniques including evidence theory for large-scale structures. Comput Struct 82(13–14):1101–1112. https://doi.org/10.1016/j.compstruc.2004.03.014

Ball DJ, Watt J (2013) Further thoughts on the utility of risk matrices. Risk Anal 33(11):2068–2078. https://doi.org/10.1111/risa.12057

Caron F (2013) Quantitative analysis of project risks. In *Managing the Continuum: Certainty, Uncertainty, Unpredictability in Large Engineering Projects* (Issue 9788847052437, pp. 75–80). Springer, Milano. https://doi.org/10.1007/978-88-470-5244-4_14

Caron F, Fumagalli M, Rigamonti A (2007) Engineering and contracting projects: A value at risk based approach to portfolio balancing. Int J Proj Manag 25(6):569–578. https://doi.org/10.1016/j.ijproman.2007.01.016

Chapman CB (1997) Project risk analysis and management– PRAM the generic process. Int J Proj Manag 15(5):273–281. https://doi.org/10.1016/S0263-7863(96)00079-8

Chapman CB, Ward S (2003) *Project Risk Management: Processes, Techniques and Insights* (John Wiley and Sons, Ed.; 2nd ed.). Chichester

Chen P-H, Peng T-T (2018) Value-at-risk model analysis of Taiwanese high-tech facility construction. J Manag Eng, *34*(2). https://doi.org/10.1061/(asce)me.1943-5479.0000585

Cox LA (2008) What's wrong with risk matrices? Risk Anal 28(2):497–512. https://doi.org/10.1111/j.1539-6924.2008.01030.x

Cox LA, Babayev D, Huber W (2005) Some limitations of qualitative risk rating systems. Risk Anal 25(3):651–662. https://doi.org/10.1111/j.1539-6924.2005.00615.x

Creemers S, Demeulemeester E, Van de Vonder S (2014) A new approach for quantitative risk analysis. Ann Oper Res 213(1):27–65. https://doi.org/10.1007/s10479-013-1355-y

Curto D, Acebes F, González-Varona JM, Poza D (2022) Impact of aleatoric, stochastic and epistemic uncertainties on project cost contingency reserves. Int J Prod Econ 253(Nov):108626. https://doi.org/10.1016/j.ijpe.2022.108626

Damnjanovic I, Reinschmidt KF (2020) *Data Analytics for Engineering and Construction Project Risk Management*. Springer International Publishing

Duijm NJ (2015) Recommendations on the use and design of risk matrices. Saf Sci 76:21–31. https://doi.org/10.1016/j.ssci.2015.02.014

Eldosouky IA, Ibrahim AH, Mohammed HED (2014) Management of construction cost contingency covering upside and downside risks. Alex Eng J 53(4):863–881. https://doi.org/10.1016/j.aej.2014.09.008

Elms DG (2004) Structural safety: Issues and progress. Prog Struct Eng Mater 6:116–126. https://doi.org/10.1002/pse.176

European Commission. (2023) *Project Management Methodology. Guide 3.1* (European Union, Ed.). Publications Office of the European Union

Frank M (1999) Treatment of uncertainties in space nuclear risk assessment with examples from Cassini mission implications. Reliab Eng Syst Safe 66:203–221. https://doi.org/10.1016/S0951-8320(99)00002-2

Gatti S, Rigamonti A, Saita F, Senati M (2007) Measuring value-at-risk in project finance transactions. Eur Financ Manag 13(1):135–158. https://doi.org/10.1111/j.1468-036X.2006.00288.x

Giot P, Laurent S (2003) Market risk in commodity markets: a VaR approach. Energy Econ 25:435–457. https://doi.org/10.1016/S0140-9883(03)00052-5

Goerlandt F, Reniers G (2016) On the assessment of uncertainty in risk diagrams. Saf Sci 84:67–77. https://doi.org/10.1016/j.ssci.2015.12.001

Helton JC, Johnson JD, Oberkampf WL, Sallaberry CJ (2006) Sensitivity analysis in conjunction with evidence theory representations of epistemic uncertainty. Reliab Eng Syst Saf 91(10–11):1414–1434. https://doi.org/10.1016/j.ress.2005.11.055

Hillson D (2014) How to manage the risks you didn't know you were taking. Paper presented at PMI® Global Congress 2014—North America, Phoenix, AZ. Newtown Square, PA: Project Management Institute

Hillson D, Simon P (2020) *Practical Project Risk Management. THE ATOM METHODOLOGY* (Third Edit, Issue 1). Berrett-Koehler Publishers, Inc

Hulett DT (2012) *Acumen Risk For Schedule Risk Analysis - A User's Perspective*. White Paper. https://info.deltek.com/acumen-risk-for-schedule-risk-analysis

International Organization for Standardization. (2018). *ISO 31000:2018 Risk management – Guidelines* (Vol. 2)

International Organization for Standardization. (2019). *ISO/IEC 31010:2019 Risk management - Risk assessment techniques*

International Project Management Association. (2015). Individual Competence Baseline for Project, Programme & Portfolio Management. Version 4.0. In *International Project Management Association* (Vol. 4). https://doi.org/10.1002/ejoc.201200111

Joukar A, Nahmens I (2016) Estimation of the Escalation Factor in Construction Projects Using Value at Risk. *Construction Research Congress*, 2351–2359. https://doi.org/10.1061/9780784479827.234

Kerzner H (2022) *Project Management. A Systems Approach to Planning, Scheduling, and Controlling* (Inc. John Wiley & Sons, Ed.; 13th Editi)

Koulinas GK, Demesouka OE, Sidas KA, Koulouriotis DE (2021) A topsis—risk matrix and Monte Carlo expert system for risk assessment in engineering projects. Sustainability 13(20):1–14. https://doi.org/10.3390/su132011277

Krisper M (2021) *Problems with Risk Matrices Using Ordinal Scales*. https://doi.org/10.48550/arXiv.2103.05440

Kuester K, Mittnik S, Paolella MS (2006) Value-at-risk prediction: A comparison of alternative strategies. J Financ Econ 4(1):53–89. https://doi.org/10.1093/jjfinec/nbj002

Kwon H, Kang CW (2019) Improving project budget estimation accuracy and precision by analyzing reserves for both identified and unidentified risks. Proj Manag J 50(1):86–100. https://doi.org/10.1177/8756972818810963

Lemmens SMP, Lopes van Balen VA, Röselaers YCM, Scheepers HCJ, Spaanderman MEA (2022) The risk matrix approach: a helpful tool weighing probability and impact when deciding on preventive and diagnostic interventions. BMC Health Serv Res 22(1):1–11. https://doi.org/10.1186/s12913-022-07484-7

Levine ES (2012) Improving risk matrices: The advantages of logarithmically scaled axes. J Risk Res 15(2):209–222. https://doi.org/10.1080/13669877.2011.634514

Li J, Bao C, Wu D (2018) How to design rating schemes of risk matrices: a sequential updating approach. Risk Anal 38(1):99–117. https://doi.org/10.1111/risa.12810

Lorance RB, Wendling RV (2001) Basic techniques for analyzing and presentation of cost risk analysis. Cost Eng 43(6):25–31

Markowski AS, Mannan MS (2008) Fuzzy risk matrix. J Hazard Mater 159(1):152–157. https://doi.org/10.1016/j.jhazmat.2008.03.055

Menge DNL, MacPherson AC, Bytnerowicz TA et al. (2018) Logarithmic scales in ecological data presentation may cause misinterpretation. Nat Ecol Evol 2:1393–1402. https://doi.org/10.1038/s41559-018-0610-7

Monat JP, Doremus S (2020) An improved alternative to heat map risk matrices for project risk prioritization. J Mod Proj Manag 7(4):214–228. https://doi.org/10.19255/JMPM02210

Naderpour H, Kheyroddin A, Mortazavi S (2019) Risk assessment in bridge construction projects in Iran using Monte Carlo simulation technique. Pract Period Struct Des Constr 24(4):1–11. https://doi.org/10.1061/(asce)sc.1943-5576.0000450

Ni H, Chen A, Chen N (2010) Some extensions on risk matrix approach. Saf Sci 48(10):1269–1278. https://doi.org/10.1016/j.ssci.2010.04.005

Peace C (2017) The risk matrix: Uncertain results? Policy Pract Health Saf 15(2):131–144. https://doi.org/10.1080/14773996.2017.1348571

Project Management Institute. (2009) *Practice Standard for Project Risk Management*. Project Management Institute, Inc

Project Management Institute. (2017) *A Guide to the Project Management Body of Knowledge: PMBoK(R) Guide. Sixth Edition* (6th ed.). Project Management Institute Inc

Project Management Institute. (2019) *The standard for Risk Management in Portfolios, Programs and Projects*. Project Management Institute, Inc

Project Management Institute. (2021) *A Guide to the Project Management Body of Knowledge: PMBoK(R) Guide. Seventh Edition* (7th ed.). Project Management Institute, Inc

Proto R, Recchia G, Dryhurst S, Freeman ALJ (2023) Do colored cells in risk matrices affect decision-making and risk perception? Insights from randomized controlled studies. *Risk Analysis*, 1–15. https://doi.org/10.1111/risa.14091

Qazi A, Dikmen I (2021) From risk matrices to risk networks in construction projects. IEEE Trans Eng Manag 68(5):1449–1460. https://doi.org/10.1109/TEM.2019.2907787

Qazi A, Shamayleh A, El-Sayegh S, Formaneck S (2021) Prioritizing risks in sustainable construction projects using a risk matrix-based Monte Carlo Simulation approach. Sustain Cities Soc 65(Aug):102576. https://doi.org/10.1016/j.scs.2020.102576

Qazi A, Simsekler MCE (2021) Risk assessment of construction projects using Monte Carlo simulation. Int J Manag Proj Bus 14(5):1202–1218. https://doi.org/10.1108/IJMPB-03-2020-0097

Rehacek P (2017) Risk management standards for project management. Int J Adv Appl Sci 4(6):1–13. https://doi.org/10.21833/ijaas.2017.06.001

Rezaei F, Najafi AA, Ramezanian R (2020) Mean-conditional value at risk model for the stochastic project scheduling problem. Comput Ind Eng 142(Jul):106356. https://doi.org/10.1016/j.cie.2020.106356

Ruan X, Yin Z, Frangopol DM (2015) Risk Matrix integrating risk attitudes based on utility theory. Risk Anal 35(8):1437–1447. https://doi.org/10.1111/risa.12400

Sarykalin S, Serraino G, Uryasev S (2008) Value-at-risk vs. conditional value-at-risk in risk management and optimization. *State-of-the-Art Decision-Making Tools in the Information-Intensive Age, October 2023*, 270–294. https://doi.org/10.1287/educ.1080.0052

Simon P, Hillson D, Newland K (1997) *PRAM Project Risk Analysis and Management Guide* (P. Simon, D. Hillson, & K. Newland, Eds.). Association for Project Management

Sutherland H, Recchia G, Dryhurst S, Freeman ALJ (2022) How people understand risk matrices, and how matrix design can improve their use: findings from randomized controlled studies. Risk Anal 42(5):1023–1041. https://doi.org/10.1111/risa.13822

Talbot, J (2014). *What's right with risk matrices? An great tool for risk managers…* 31000risk. https://31000risk.wordpress.com/article/what-s-right-with-risk-matrices-3dksezemjiq54-4/

Taroun A (2014) Towards a better modelling and assessment of construction risk: Insights from a literature review. Int J Proj Manag 32(1):101–115. https://doi.org/10.1016/j.ijproman.2013.03.004

The Standish Group. (2022). *Chaos report*. https://standishgroup.myshopify.com/collections/all

Thomas P, Bratvold RB, Bickel JE (2014) The risk of using risk matrices. SPE Econ Manag 6(2):56–66. https://doi.org/10.2118/166269-pa

Tong R, Cheng M, Zhang L, Liu M, Yang X, Li X, Yin W (2018) The construction dust-induced occupational health risk using Monte-Carlo simulation. J Clean Prod 184:598–608. https://doi.org/10.1016/j.jclepro.2018.02.286

Traynor BA, Mahmoodian M (2019) Time and cost contingency management using Monte Carlo simulation. Aust J Civ Eng 17(1):11–18. https://doi.org/10.1080/14488353.2019.1606499

Vanhoucke, M (2018). The data-driven project manager: A statistical battle against project obstacles. In *The Data-Driven Project Manager: A Statistical Battle Against Project Obstacles*. https://doi.org/10.1007/978-1-4842-3498-3

Vatanpour S, Hrudey SE, Dinu I (2015) Can public health risk assessment using risk matrices be misleading? Int J Environ Res Public Health 12(8):9575–9588. https://doi.org/10.3390/ijerph120809575

Vose, D (2008). *Risk Analysis: a Quantitative Guide (3rd ed.)*. Wiley
Votto R, Lee Ho L, Berssaneti F (2020a) Applying and assessing performance of earned duration management control charts for EPC project duration monitoring. J Constr Eng Manag 146(3):1–13. https://doi.org/10.1061/(ASCE)CO.1943-7862.0001765
Votto R, Lee Ho L, Berssaneti F (2020b) Multivariate control charts using earned value and earned duration management observations to monitor project performance. Comput Ind Eng 148(Sept):106691. https://doi.org/10.1016/j.cie.2020.106691
Ward S (1999) Assessing and managing important risks. Int J Proj Manag 17(6):331–336. https://doi.org/10.1016/S0263-7863(98)00051-9

## Acknowledgements
This research has been partially funded by the Regional Government of Castile and Leon (Spain) and the European Regional Development Fund (ERDF, FEDER) with grant VA180P20.

## Author contributions
FA developed the conceptualisation and the methodology. JMG contributed to the literature review and interpretations of the results for the manuscript. FA and JP collected the experimental data and developed all the analyses and simulations. AL supervised the project. FA and JP wrote the original draft, while AL and JMG conducted the review and editing. All authors have read and agreed to the published version of the manuscript.

## Competing interests
The authors declare no competing interests.

## Ethical approval
Ethical approval was not required as the study did not involve human participants.

## Informed consent
No human subjects are involved in this study.

## Additional information
**Supplementary information** The online version contains supplementary material available at https://doi.org/10.1057/s41599-024-03180-5.

**Correspondence** and requests for materials should be addressed to F. Acebes.

**Reprints and permission information** is available at http://www.nature.com/reprints

**Publisher's note** Springer Nature remains neutral with regard to jurisdictional claims in published maps and institutional affiliations.